**New copulas based on general partitions-of-unity and their applications to risk management**

Dietmar Pfeifer[1)], Hervé Awoumlac Tsatedem†[1)], Andreas Mändle[1)], and Côme Girschig[2)]

Carl von Ossietzky Universität Oldenburg, Germany[1)] and
École Nationale des Ponts et Chaussées, Paris, France[2)]

**July 8, 2016**

**Abstract:** We construct new multivariate copulas on the basis of a generalized infinite partition-of-unity approach. This approach allows - in contrast to finite partition-of-unity copulas - for tail-dependence as well as for asymmetry. A possibility of fitting such copulas to real data from quantitative risk management is also pointed out.

**1. Introduction**

The theory of copulas and their applications has gained much interest in the recent years, especially in the field of quantitative risk management, insurance and finance (see e.g. MCNEIL, FREY AND EMBRECHTS (2005) or RANK (2006)). While classical approaches like elliptically contoured copulas and Archimedean copulas are widely explored, other approaches concentrate on non-standard, non-symmetric or data-driven copula constructions (see e.g. LAUTERBACH AND PFEIFER (2015), LAUTERBACH (2014), COTTIN AND PFEIFER (2014) or JAWORSKI, DURANTE AND HÄRDLE (2013) and the papers therein for a survey, especially the contributions related to vine copulas). Statistical and computational aspects of copulas have also been investigated in more detail recently (see e.g. BLUMENTRITT (2012) and MAI AND SCHERER (2012)). In this paper, we want to focus on a particular class of copulas and their generalizations, the so called partition-of-unity copulas (see e.g. LI, MIKUSIŃSKI AND TAYLOR (1998) or KULPA (1999)). Whereas in the usual approach, only finite partitions-of-unity are considered, which do not allow for a modelling of tail-dependence, we extend this concept to infinite partitions-of-unity, which allows for tail-dependence as well as for asymmetry, and which can also be used to fit given data to a more realistic copula model. Our investigations resemble in some sense more recent approaches such as YANG ET AL. (2015), GONZÁLEZ-BARRIOS AND HERNÁNDEZ-CEDILLO (2013), ZHENG ET AL. (2011), HUMMEL AND MÄRKERT (2011), or GHOSH AND HENDERSON (2009). Whereas in these papers, local modifications of known standard copulas are considered in order to obtain tail dependence or asymmetries, we focus on a closed form representation of completely new copula densities which allows for easy Monte Carlo simulations as well as a data driven modelling of tail dependence and asymmetries. This approach is not restricted to two dimensions in general, but can likewise be used in arbitrary dimensions. However, in order to illustrate our results, we will give examples in the bivariate case only.

To facilitate the readability of the paper, all elaborate proofs are given in an appendix.

## 2. Main Results

Let $\mathbb{Z}^+ = \{0,1,2,3,\cdots\}$ denote the set of non-negative integers and suppose that $\{\varphi_i(u)\}_{i\in\mathbb{Z}^+}$ and $\{\psi_j(v)\}_{j\in\mathbb{Z}^+}$ are non-negative maps defined on the interval $(0,1)$ each such that

$$\sum_{i=0}^{\infty} \varphi_i(u) = \sum_{j=0}^{\infty} \psi_j(v) = 1 \qquad (2.1)$$

and

$$\int_0^1 \varphi_i(u)\,du = \alpha_i > 0, \ \int_0^1 \psi_j(v)\,dv = \beta_j > 0 \ \text{ for } i,j \in \mathbb{Z}^+. \qquad (2.2)$$

The maps $\varphi_i(u)$ and $\psi_j(v)$ can be thought of as representing discrete distributions over the non-negative integers $\mathbb{Z}^+$ with parameters $u$ and $v$, resp. The sequences $\{\alpha_i\}_{i\in\mathbb{Z}^+}$ and $\{\beta_j\}_{j\in\mathbb{Z}^+}$ then represent the probabilities of the corresponding mixed distributions each.

Let further $\{p_{ij}\}_{i,j\in\mathbb{Z}^+}$ represent the probabilities of an arbitrary discrete bivariate distribution over $\mathbb{Z}^+ \times \mathbb{Z}^+$ with marginal distributions given by $p_{i\cdot} = \sum_{j=0}^{\infty} p_{ij} = \alpha_i$ and $p_{\cdot j} = \sum_{i=0}^{\infty} p_{ij} = \beta_j$ for $i,j \in \mathbb{Z}^+$. Then

$$c(u,v) := \sum_{i=0}^{\infty}\sum_{j=0}^{\infty} \frac{p_{ij}}{\alpha_i \beta_j} \varphi_i(u)\psi_j(v), \ u,v \in (0,1) \qquad (2.3)$$

defines the density of a bivariate copula, called *generalized partition-of-unity copula*. The fact that $c$ in fact is the density of a bivariate copula can be seen as follows:

$$\int_0^1 c(u,v)\,dv = \sum_{i=0}^{\infty}\sum_{j=0}^{\infty} \frac{p_{ij}}{\alpha_i \beta_j} \varphi_i(u) \int_0^1 \psi_j(v)\,dv = \sum_{i=0}^{\infty}\sum_{j=0}^{\infty} \frac{p_{ij}}{\alpha_i \beta_j} \beta_j \varphi_i(u)$$

$$= \sum_{i=0}^{\infty}\sum_{j=0}^{\infty} \frac{p_{ij}}{\alpha_i} \varphi_i(u) = \sum_{i=0}^{\infty} \frac{\varphi_i(u)}{\alpha_i} \sum_{j=0}^{\infty} p_{ij} = \sum_{i=0}^{\infty} \frac{\varphi_i(u)}{\alpha_i} \alpha_i = \sum_{i=0}^{\infty} \varphi_i(u) = 1, \qquad (2.4)$$

likewise for $\int_0^1 c(u,v)\,du$.

Note that from a „dual" point of view, we can rewrite (2.3) as

$$c(u,v) = \sum_{i=0}^{\infty}\sum_{j=0}^{\infty} p_{ij} f_i(u) g_j(v), \ u,v \in (0,1) \qquad (2.5)$$

where $f_i(\bullet) = \dfrac{\varphi_i(\bullet)}{\alpha_i}$, $g_j(\bullet) = \dfrac{\psi_j(\bullet)}{\beta_j}$, $i, j \in \mathbb{Z}^+$ denote the Lebesgue densities induced by $\{\varphi_i(u)\}_{i \in \mathbb{Z}^+}$ and $\{\psi_j(v)\}_{j \in \mathbb{Z}^+}$. This means that the copula density $c$ can also be seen as an appropriate mixture of product densities, which possibly allows for a simple way for a stochastic simulation.

An extension of this approach to $d$ dimensions with $d > 2$ is obvious: assume that $\{\varphi_{ki}(u)\}_{i \in \mathbb{Z}^+}$ for $k = 1, \cdots, d$ represent discrete probabilities with

$$\sum_{i=0}^{\infty} \varphi_{ki}(u) = 1 \text{ for } u \in (0,1) \tag{2.6}$$

and

$$\int_0^1 \varphi_{ki}(u)\, du = \alpha_{ki} > 0 \text{ for } i \in \mathbb{Z}^+. \tag{2.7}$$

Let further $\{p_{\mathbf{i}}\}_{\mathbf{i} \in \mathbb{Z}^{+d}}$ represent the distribution of an arbitrary discrete $d$-dimensional random vector $\mathbf{Z}$ over $\mathbb{Z}^{+d}$ where, for simplicity, we write $\mathbf{i} = (i_1, \cdots, i_d)$, i.e.

$$P(\mathbf{Z} = \mathbf{i}) = p_{\mathbf{i}}, \ \mathbf{i} \in \mathbb{Z}^{+d}. \tag{2.8}$$

Suppose further that for the marginal distributions, there holds

$$P(Z_k = i) = \alpha_{ki}, \ i \in \mathbb{Z}^+, \ k = 1, \cdots, d. \tag{2.9}$$

Then

$$c(\mathbf{u}) := \sum_{\mathbf{i} \in \mathbb{Z}^{+d}} \dfrac{p_{\mathbf{i}}}{\prod_{k=1}^{d} \alpha_{k,i_k}} \prod_{k=1}^{d} \varphi_{k,i_k}(u_k), \ \mathbf{u} = (u_1, \cdots, u_d) \in (0,1)^d \tag{2.10}$$

defines the density of a $d$-variate copula, which is also called *generalized partition-of-unity copula*.

Alternatively, we can rewrite (2.10) again as

$$c(\mathbf{u}) = \sum_{\mathbf{i} \in \mathbb{Z}^{+d}} p_{\mathbf{i}} \prod_{k=1}^{d} f_{k,i_k}(u_k), \ \mathbf{u} = (u_1, \cdots, u_d) \in (0,1)^d \tag{2.11}$$

where the $f_{ki}(\bullet) = \dfrac{\varphi_{ki}(\bullet)}{\alpha_{ki}}$, $i \in \mathbb{Z}^+$, $k = 1, \cdots, d$ denote the Lebesgue densities induced by the $\{\varphi_{ki}(u)\}_{i \in \mathbb{Z}^+}$.

## 3. The symmetric case (diagonal dominance)

For simplicity, we restrict ourselves to the two-dimensional case in the sequel. The generalization to higher dimensions is obvious.

Let $\varphi_i = \psi_i$ for $i \in \mathbb{Z}^+$ and $\int_0^1 \varphi_i(u)\,du = \alpha_i > 0$. Define

$$p_{ij} := \begin{cases} \alpha_i, & \text{if } i = j \\ 0, & \text{otherwise.} \end{cases} \quad (3.1)$$

Then

$$c(u,v) := \sum_{i=0}^{\infty} \frac{\varphi_i(u)\varphi_i(v)}{\alpha_i} = \sum_{i=0}^{\infty} \alpha_i f_i(u) f_i(v), \ u,v \in (0,1) \quad (3.2)$$

defines the density of a bivariate copula, called *generalized partition-of-unity copula* with *diagonal dominance*.

**Example 1** (binomial distributions - Bernstein copula)**.** Consider, for a fixed integer $m \geq 2$, the family of binomial distributions given by their point masses

$$\varphi_{m,i}(u) = \begin{cases} \binom{m-1}{i} u^i (1-u)^{m-1-i}, & i = 0, \cdots, m-1 \\ 0, & i \geq m. \end{cases} \quad (3.3)$$

Here we have, for $i = 0, \cdots, m-1$,

$$\alpha_{m,i} = \int_0^1 \varphi_{m,i}(u)\,du = \binom{m-1}{i} \int_0^1 u^i (1-u)^{m-1-i}\,du$$
$$= \frac{(m-1)!}{i!(m-1-i)!} \cdot \frac{\Gamma(i+1)\Gamma(m-i)}{\Gamma(m+1)} = \frac{(m-1)!}{i!(m-1-i)!} \cdot \frac{i!(m-1-i)!}{m!} = \frac{1}{m} \quad (3.4)$$

and hence

$$c_m(u,v) = m \sum_{i=0}^{m-1} \binom{m-1}{i}^2 (uv)^i \big((1-u)(1-v)\big)^{m-1-i}, \ u,v \in (0,1) \quad (3.5)$$

which corresponds to the density of a particular Bernstein copula (see e.g. COTTIN AND PFEIFER (2014), Theorem 2.1). Especially, for $m = 2$, we obtain

$$c_2(u,v) = 4uv - 2u - 2v + 2, \ u,v \in (0,1). \quad (3.6)$$

The corresponding copula $C_2$ is given by

$$C_2(x,y) = \int_0^x \int_0^y c_2(u,v)\,dv\,du = xy + xy(1-x)(1-y), \quad x,y \in (0,1) \tag{3.7}$$

and belongs to the so called *Farlie-Gumbel-Morgenstern* family (cf. e.g. NELSEN (2006), p. 77). For general $m > 1$, relation (3.5) represents the density of a copula with polynomial sections of degree $m$ in both variables (cf. NELSEN (2006), chapter 3.2.5). The following graphs show some of these densities for different values of $m$.

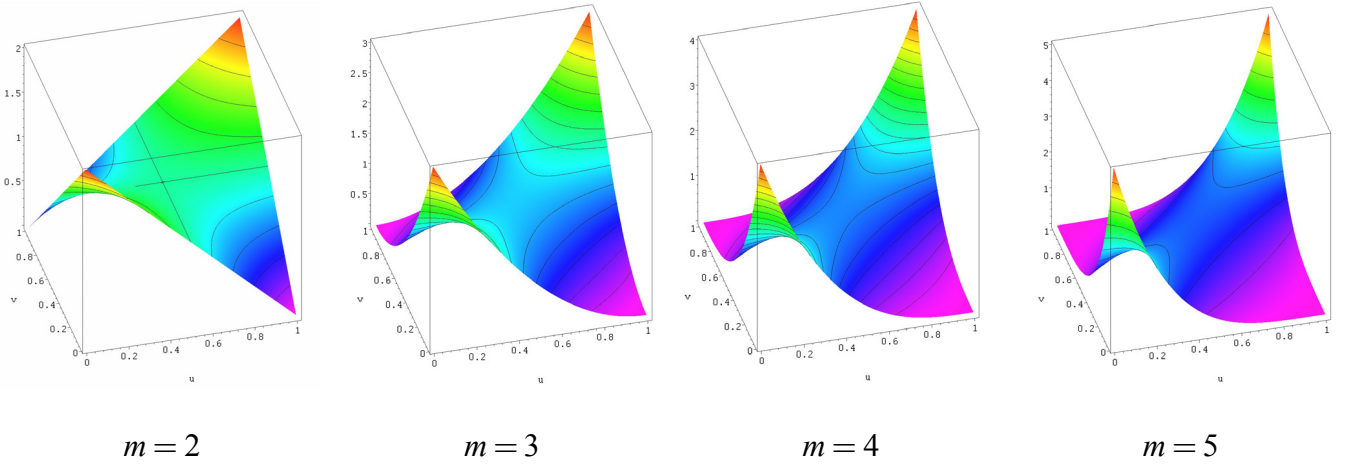

$m = 2$  $\quad\quad\quad\quad\quad$  $m = 3$  $\quad\quad\quad\quad\quad$  $m = 4$  $\quad\quad\quad\quad\quad$  $m = 5$

Clearly, all those densities are bounded by the constant $m$, hence the coefficients $\lambda_U$ and $\lambda_L$ of upper and lower tail dependence are zero:

$$\lambda_U = \lim_{t\uparrow 1} \frac{\int_t^1 \int_t^1 c_m(u,v)\,du\,dv}{1-t} \le \lim_{t\uparrow 1} \frac{m(1-t)^2}{1-t} = 0 \text{ and } \lambda_L = \lim_{t\downarrow 0} \frac{\int_0^t \int_0^t c_m(u,v)\,du\,dv}{t} \le \lim_{t\downarrow 0} \frac{mt^2}{t} = 0. \tag{3.8}$$

**Example 2** (negative binomial distributions). Consider, for fixed $\beta > 0$, the family of negative binomial distributions given by their point masses

$$\varphi_{\beta,i}(u) = \binom{\beta+i-1}{i}(1-u)^\beta u^i, \quad i \in \mathbb{Z}^+. \tag{3.9}$$

Here we have, for $i \in \mathbb{Z}^+$,

$$\alpha_{\beta,i} = \int_0^1 \varphi_{\beta,i}(u)\,du = \binom{\beta+i-1}{i}\int_0^1 u^i(1-u)^\beta\,du = \frac{\Gamma(\beta+i)}{i!\,\Gamma(\beta)} \cdot \frac{\Gamma(i+1)\Gamma(\beta+1)}{\Gamma(\beta+i+2)} = \frac{\beta}{(\beta+i)(\beta+i+1)} \tag{3.10}$$

and hence

$$c_\beta(u,v) = \frac{((1-u)(1-v))^\beta}{\beta} \sum_{i=0}^{\infty} (\beta+i)(\beta+i+1)\binom{\beta+i-1}{i}^2 (uv)^i$$

$$= (\beta+1)((1-u)(1-v))^\beta \sum_{i=0}^{\infty} \binom{\beta+i-1}{i}\binom{\beta+i+1}{i}(uv)^i, \ u,v \in (0,1). \tag{3.11}$$

For integer choices of $\beta$, this expression can be explicitly evaluated as a finite sum, as can be seen from the following result.

**Lemma 1.** For $\beta \in \mathbb{N}$, there holds

$$c_\beta(u,v) = (\beta+1)\frac{((1-u)(1-v))^\beta}{(1-uv)^{2\beta+1}} \sum_{i=0}^{\beta-1} \binom{\beta-1}{i}\binom{\beta+1}{i}(uv)^i, \ u,v \in (0,1). \tag{3.12}$$

To give an illustration of Lemma 1, we show an exemplary table for $\beta = 1,\cdots,6$, likewise for the corresponding copula $C_\beta(x,y) = \int_0^x \int_0^y c_\beta(u,v)\,dv\,du, \ x,y \in (0,1)$.

| $\beta$ | $c_\beta(u,v), \ u,v \in (0,1)$ |
|---|---|
| 1 | $2\dfrac{(1-u)(1-v)}{(1-uv)^3}$ |
| 2 | $3\dfrac{(1+3uv)(1-u)^2(1-v)^2}{(1-uv)^5}$ |
| 3 | $4\dfrac{(1+8uv+6u^2v^2)(1-u)^3(1-v)^3}{(1-uv)^7}$ |
| 4 | $5\dfrac{(1+15uv+30u^2v^2+10u^3v^3)(1-u)^4(1-v)^4}{(1-uv)^9}$ |
| 5 | $6\dfrac{(1+24uv+90u^2v^2+80u^3v^3+15u^4v^4)(1-u)^5(1-v)^5}{(1-uv)^{11}}$ |
| 6 | $7\dfrac{(1+35uv+210u^2v^2+350u^3v^3+175u^4v^4+21u^5v^5)(1-u)^6(1-v)^6}{(1-uv)^{13}}$ |

| $\beta$ | $C_\beta(x,y)$, $x,y \in (0,1)$ |
|---|---|
| 1 | $xy\dfrac{(2-x-y)}{1-xy}$ |
| 2 | $xy\dfrac{\left(3-3x-3y+x^2+y^2+3x^2y^2-x^2y^3-x^3y^2\right)}{(1-xy)^3}$ |
| 3 | $\dfrac{xy}{(1-xy)^5}\big(4-6x-6y+4x^2+4xy+4y^2+4x^3y+24x^2y^2+4xy^3-x^3-6x^2y-6xy^2-y^3-\cdots$ $\cdots-x^5y^4-x^4y^5+4x^4y^4-x^4y^3-x^3y^4+4x^4y^2+4x^3y^3+4x^2y^4-x^4y-16x^3y^2-16x^2y^3-xy^4\big)$ |

The following graphs show the negative binomial copula densities $c_\beta$ for $\beta = 1, \cdots, 6$.

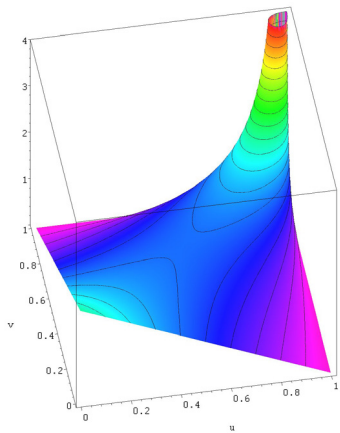

$\beta = 1$

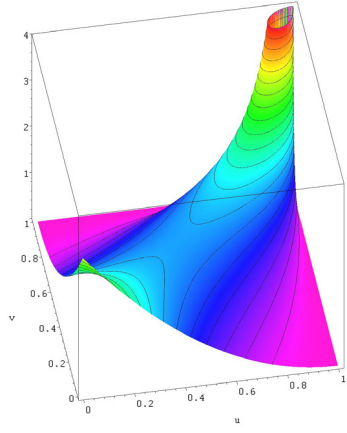

$\beta = 2$

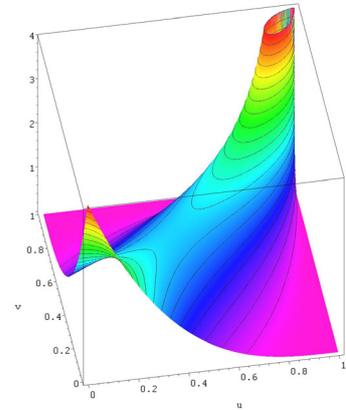

$\beta = 3$

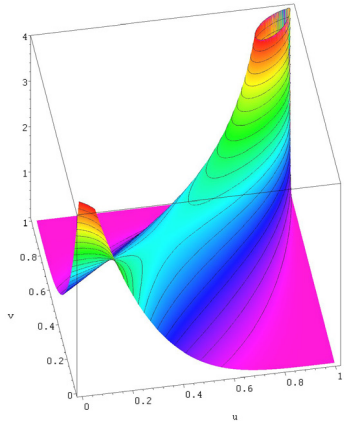

$\beta = 4$

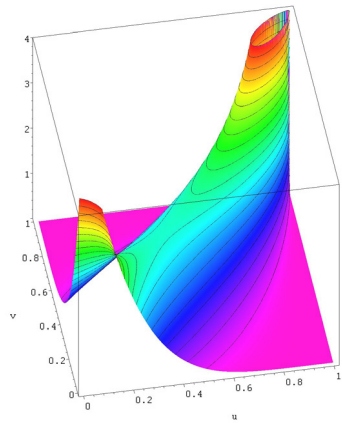

$\beta = 5$

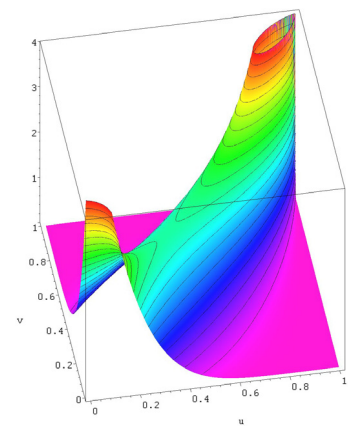

$\beta = 6$

Negative binomial copulas typically show an upper tail dependence, as can be seen from the following exemplary table.

| $\beta$ | 1 | 2 | 3 | 4 | 5 | 6 | 7 | 8 | 9 | 10 |
|---|---|---|---|---|---|---|---|---|---|---|
| $\lambda_U(\beta)$ | $\dfrac{1}{2}$ | $\dfrac{5}{8}$ | $\dfrac{11}{16}$ | $\dfrac{93}{128}$ | $\dfrac{193}{256}$ | $\dfrac{793}{1024}$ | $\dfrac{1619}{2048}$ | $\dfrac{26333}{32768}$ | $\dfrac{53381}{65536}$ | $\dfrac{215955}{262144}$ |

A closed formula for the tail dependence coefficients for integer values of $\beta$ is given in the following result.

**Lemma 2.** For $\beta \in \mathbb{N}$, there holds

$$\lambda_U(\beta) = \lim_{t \uparrow 1} \frac{\int_t^1 \int_t^1 c_\beta(u,v)\,du\,dv}{1-t} = \frac{2\Gamma(2\beta)}{\Gamma^2(\beta)} \cdot \int_0^1 \int_0^1 \frac{x^\beta y^\beta}{(x+y)^{2\beta+1}}\,dx\,dy = \frac{4\Gamma(2\beta)}{\Gamma^2(\beta)} \cdot \int_0^{1/2} u^\beta (1-u)^{\beta-1}\,du$$

$$= 1 - \frac{\binom{2\beta}{\beta}}{4^\beta} \sim 1 - \frac{1}{\sqrt{\pi\beta}} \quad \text{for large } \beta. \tag{3.13}$$

Note that the sequence $4^\beta \lambda_U(\beta) = 4^\beta - \binom{2\beta}{\beta}$ is related to certain combinatorial graph problems, see LISK-OVETZ AND WALSH (2006), Table 4, p.385. The authors remark in their paper: "The latter [sequence] is also known as the enumerator of cycles of objects, where the individual objects are enumerated by the Catalan numbers."

Note that relation (3.13) also implies that $\lim_{\beta \to \infty} \lambda_U(\beta) = 1$.

**Example 3** (Poisson distributions). Consider the family of Poisson distributions given by their point masses

$$\varphi_{\gamma,i}(u) = (1-u)^\gamma \frac{\gamma^i L(u)^i}{i!}, \quad i \in \mathbb{Z}^+ \tag{3.14}$$

where $L(u) = -\ln(1-u) > 0$, $u \in (0,1)$ and $\gamma > 0$. Here we get, for $i \in \mathbb{Z}^+$, with the substitutions $z = L(u)$ and $y = (1+\gamma)z$,

$$\alpha_{\gamma,i} = \int_0^1 \varphi_{\gamma,i}(u)\,du = \int_0^1 (1-u)^\gamma \frac{\gamma^i L(u)^i}{i!}\,du = \int_0^\infty \frac{\gamma^i z^i}{i!} e^{-(1+\gamma)z}\,dz$$

$$= \frac{\gamma^i}{(1+\gamma)^{i+1}} \int_0^\infty \frac{y^i}{i!} e^{-y}\,dy = \frac{\gamma^i}{(1+\gamma)^{i+1}} = \left(\frac{\gamma}{1+\gamma}\right)^i \left(1 - \frac{\gamma}{1+\gamma}\right), \tag{3.15}$$

indicating that the $\alpha_{\gamma,i}$ correspond to a geometric distribution with mean $\gamma$, and hence

$$c_\gamma(u,v) = (1+\gamma)(1-u)^\gamma (1-v)^\gamma \sum_{i=0}^\infty \frac{(\gamma(1+\gamma)\ln(1-u)\ln(1-v))^i}{i!^2}, \quad u,v \in (0,1). \tag{3.16}$$

The following graphs show some of these copula densities for different choices of $\gamma$.

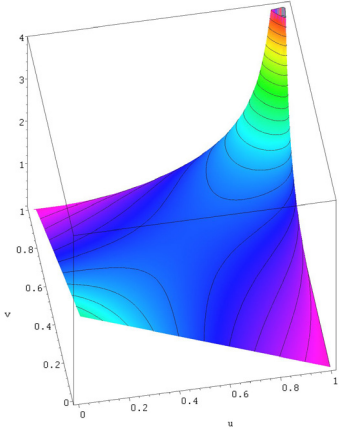 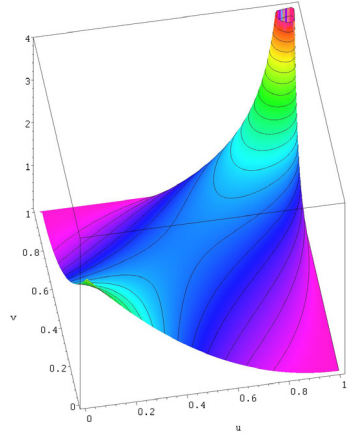 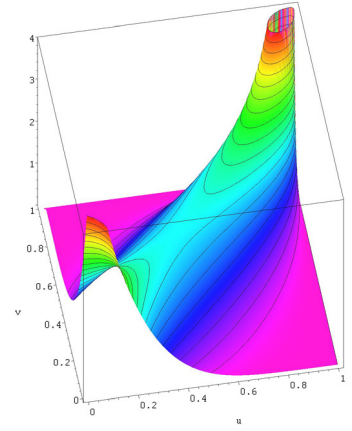

$\gamma = 1$  $\gamma = 2$  $\gamma = 5$

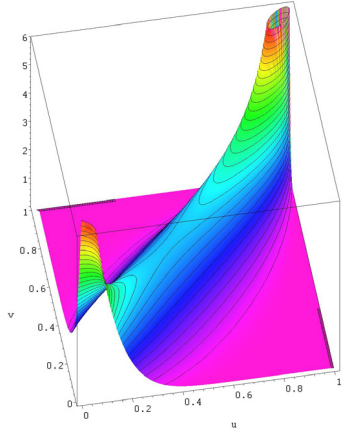 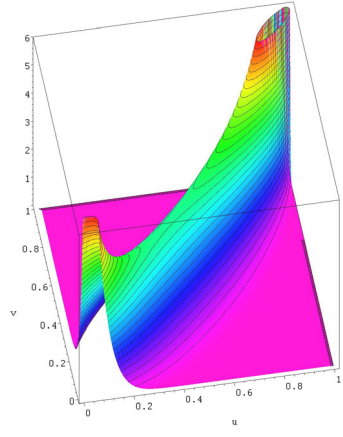 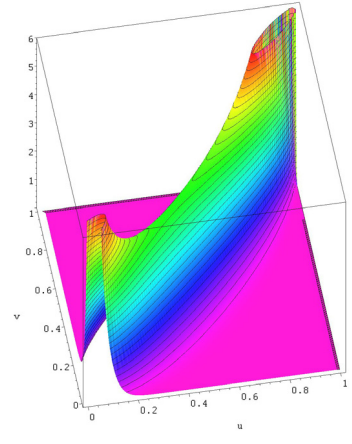

$\gamma = 10$  $\gamma = 20$  $\gamma = 30$

The corresponding copula $C$ cannot be calculated explicitly. However, in contrast to the visual impression, the coefficient $\lambda_U(\gamma)$ of upper tail dependence is zero here for all $\gamma > 0$, although we have a singularity in the point $(1,1)$ in all cases.

For a rigorous proof, we first remark that

$$h(x,y) := \sum_{i=0}^{\infty} \frac{x^i y^i}{i!^2} \leq \left(\sum_{i=0}^{\infty} \frac{x^i}{i!}\right) \cdot \left(\sum_{i=0}^{\infty} \frac{y^i}{i!}\right) = \exp(x+y) \quad \text{for all} \quad x, y \geq 0 \tag{3.17}$$

such that, with the constant $K := \gamma - \sqrt{\gamma(1+\gamma)}$,

$$c_\gamma(u,v) = (1+\gamma)(1-u)^\gamma (1-v)^\gamma h\left(-\sqrt{\gamma(1+\gamma)} \ln(1-u), -\sqrt{\gamma(1+\gamma)} \ln(1-v)\right)$$
$$\leq 2(1-u)^K (1-v)^K, \; u,v \in (0,1). \tag{3.18}$$

This implies

$$\lambda_U(\gamma) = \lim_{t \uparrow 1} \frac{\int_t^1 \int_t^1 c_\gamma(u,v)\, du\, dv}{1-t} \leq 2 \lim_{t \uparrow 1} \frac{(1-t)^{2K+1}}{(K+1)^2} = 0, \tag{3.19}$$

as stated. (Note that $2K+1 = 1+2\gamma - 2\sqrt{\gamma(1+\gamma)} > 0$.)

**Example 4** (log series distribution). Consider the family of log series distributions given by their point masses

$$\varphi_i(u) = \frac{u^i}{i \cdot L(u)}, \quad i \in \mathbb{N} \tag{3.20}$$

where again $L(u) = -\ln(1-u)$, $u \in (0,1)$. Here we get

$$\alpha_i = \int_0^1 \varphi_i(u)\,du = \frac{1}{i}\sum_{j=1}^{i}\binom{i}{j}(-1)^{j+1}\ln(j+1) \text{ for } i \in \mathbb{N}. \tag{3.21}$$

The proof of this relation requires some more sophisticated arguments, as is shown in the sequel.

**Lemma 3.** For $c > 0$ and $n \in \mathbb{N}$, there holds

$$\int_0^\infty \frac{(1-e^{-x})^n}{x} e^{-cx}\,dx = \sum_{j=0}^{n}\binom{n}{j}(-1)^{j+1}\bigl(\ln(j+c)-\ln(c)\bigr). \tag{3.22}$$

Note that for the special case $c = 1$ we obtain, by the substitution $x = -\ln(1-u)$,

$$\beta_n := \int_0^1 \frac{u^n}{-\ln(1-u)}\,du = \int_0^\infty \frac{(1-e^{-x})^n}{x} e^{-x}\,dx = \sum_{j=1}^{n}\binom{n}{j}(-1)^{j+1}\ln(j+1). \tag{3.23}$$

Hence with $\varphi_i(u) = -\dfrac{u^i}{i \cdot \ln(1-u)}$ for $i \in \mathbb{N}$, this means

$$\alpha_i = \int_0^1 \varphi_i(u)\,du = \frac{\beta_i}{i} = \frac{1}{i}\sum_{j=1}^{i}\binom{i}{j}(-1)^{j+1}\ln(j+1) \text{ for } i \in \mathbb{N}. \tag{3.24}$$

The density of the bivariate log series copula is hence given by

$$c(u,v) = \sum_{i=1}^\infty \frac{1}{\alpha_i}\varphi_i(u)\varphi_i(v) = \frac{1}{\ln(1-u)\ln(1-v)}\sum_{i=1}^\infty \frac{(uv)^i}{i\beta_i} \text{ for } 0 < u,v < 1. \tag{3.25}$$

The following graph shows the corresponding copula density.

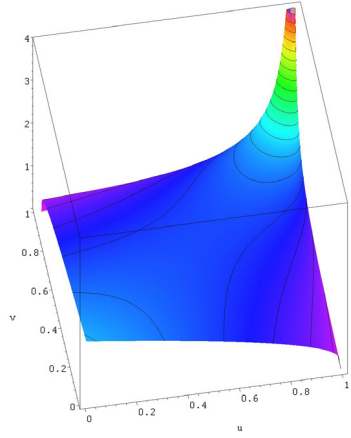

plot of $c(u,v)$

The log series copula does not have a positive tail dependence either, as in the case of the Poisson copula. The proof of this statement again requires some more sophisticated arguments. We proceed in the following steps.

**Lemma 4.** With $L(u) = -\ln(1-u)$, we have

$$\lim_{t \uparrow 1} \frac{1}{1-t} \int_t^1 \frac{L(ut)}{L(u)} du = 1. \tag{3.26}$$

**Lemma 5.** With $L(u) = -\ln(1-u)$, the $\alpha_i$ given in (3.21) and the copula density given in (3.25), it holds that

$$K(t) := \frac{1}{1-t} \int_t^1 \int_t^1 c(u,v) \, du \, dv \leq 1 - \frac{t}{1-t} \int_t^1 \frac{L(ut)}{L(u)} du \quad \text{for } 0 < t < 1, \tag{3.27}$$

which in turn implies that the log series copula has no tail dependence.

## 4. The asymmetric case

Specifying the probabilities $p_{ij}$ in a non-symmetric way we obtain asymmetric copula densities even if the maps $\varphi_i(\bullet)$ and $\psi_j(\bullet)$ are identical. A very simple approach to this problem is a specification of a suitable non-symmetric $(n+1) \times (n+1)$-matrix $M_n = [p_{ij}]_{i,j=0,\cdots,n}$ for $n \in \mathbb{Z}^+$ with

$$\sum_{k=0}^n p_{ik} = \sum_{k=0}^n p_{ki} = \alpha_i \quad \text{for } i = 0, \cdots, n \tag{4.1}$$

and

$$p_{ij} := \begin{cases} \alpha_i, & \text{if } i = j \\ 0, & \text{otherwise} \end{cases} \quad \text{for } i, j > n. \tag{4.2}$$

**Example 5** (negative binomial distributions, asymmetric case). We consider the negative binomial distributions from Example 2 with $\beta = 1$. Then $\alpha_i = \int_0^1 \varphi_{1,i}(u)\,du = \dfrac{1}{(1+i)(2+i)}$ for $i \in \mathbb{Z}^+$. With $n = 4$ and

$$M_4 := \frac{1}{60} \begin{bmatrix} 18 & 5 & 5 & 0 & 2 \\ 10 & 0 & 0 & 0 & 0 \\ 0 & 5 & 0 & 0 & 0 \\ 0 & 0 & 0 & 3 & 0 \\ 2 & 0 & 0 & 0 & 0 \end{bmatrix} \tag{4.3}$$

the conditions above are fulfilled, giving the copula density, according to (2.3),

$$c(u,v) = \sum_{i=0}^{\infty} \sum_{j=0}^{\infty} \frac{p_{ij}}{\alpha_i \alpha_j} \varphi_i(u) \varphi_j(v) = \sum_{i=0}^{n} \sum_{j=0}^{n} \frac{p_{ij}}{\alpha_i \alpha_j} \varphi_i(u) \varphi_j(v) + \sum_{k=n+1}^{\infty} \frac{1}{\alpha_k} \varphi_k(u) \varphi_k(v), \quad u, v \in (0,1), \tag{4.4}$$

or, more explicitly,

$$c(u,v) = (1-u)(1-v) \cdot \begin{pmatrix} 2 & 6u & 12u^2 & 20u^3 & 30u^4 \end{pmatrix} \cdot M_4 \cdot \begin{pmatrix} 2 \\ 6v \\ 12v^2 \\ 20v^3 \\ 30v^4 \end{pmatrix} + \dots$$

$$+ (1-u)(1-v) \cdot \sum_{k=5}^{\infty} (k+1)(k+2) u^k v^k = \frac{(1-u)(1-v)}{5(1-uv)^3} H(u,v), \quad u, v \in (0,1) \tag{4.5}$$

with the polynomial

$$H(u,v) = 150u^7v^7 - 450u^6v^6 - 10u^7v^3 + 510u^5v^5 - 10u^3v^7 - 30u^5v^4 - 10u^3v^5 + 30u^2v^6 - \dots$$
$$\dots - 300u^4v^4 + 30u^6v^2 - 5u^3v^4 + 80u^4v^3 - 30u^5v + 94u^3v^3 + 30u^2v^4 - 30uv^5 - 60u^3v^2 + \dots$$
$$\dots + 15u^2v^3 + 10u^4 + 18u^2v^2 - 30uv^3 + 10v^4 - 15uv^2 + 10v^2 - 18uv + 10u + 5v + 6. \tag{4.6}$$

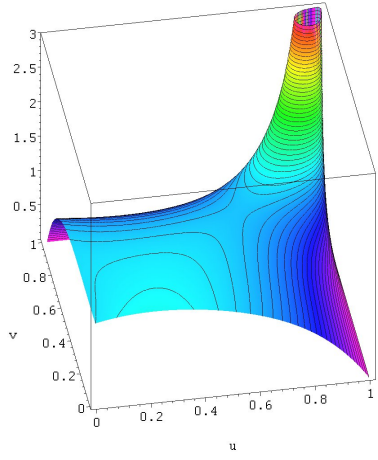 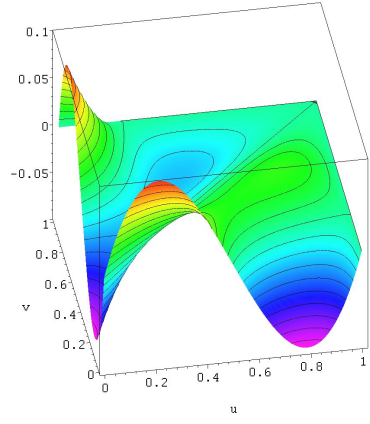

plot of $c(u,v)$          plot of $c(u,v)-c(v,u)$

The corresponding copula $C$ again has a coefficient of upper tail dependence $\lambda_U = \frac{1}{2}$ as in the symmetric case.

The following example shows an asymmetric copula composed by two different negative binomial distributions.

**Example 6.** We consider the negative binomial distributions from Example 2 with $\beta = 1$ and $\beta = 2$. Then $\alpha_i = \int_0^1 \varphi_{1,i}(u)\,du = \frac{1}{(1+i)(2+i)}$ and $\beta_j = \int_0^1 \varphi_{2,j}(v)\,dv = \frac{2}{(2+j)(3+j)} = 2\alpha_{j+1}$ for $i,j \in \mathbb{Z}^+$. Let further

$$p_{ij} = \begin{cases} \beta_j & \text{if } j = 2i \\ \beta_j & \text{if } j = 2i+1 \quad \text{for } i,j \in \mathbb{Z}^+, \\ 0 & \text{otherwise} \end{cases} \tag{4.7}$$

i.e.

$$[p_{ij}]_{i,j\in\mathbb{Z}^+} = \begin{bmatrix} \beta_0 & \beta_1 & \cdots & \cdots & \cdots & \cdots & \cdots & \cdots & \cdots \\ \cdots & \cdots & \beta_2 & \beta_3 & \cdots & \cdots & \cdots & \cdots & \cdots \\ \cdots & \cdots & \cdots & \cdots & \beta_4 & \beta_5 & \cdots & \cdots & \cdots \\ \cdots & \cdots & \cdots & \cdots & \cdots & \cdots & \beta_6 & \beta_7 & \cdots \\ \cdots & \cdots & \cdots & \cdots & \cdots & \cdots & \cdots & \cdots & \end{bmatrix} \tag{4.8}$$

where $\cdots$ stands for zero. Then $p_{i\bullet} = \sum_{j=0}^{\infty} p_{ij} = \alpha_i$ and $p_{\bullet j} = \sum_{i=0}^{\infty} p_{ij} = \beta_j$ for $i,j \in \mathbb{Z}^+$ since

$$\beta_{2i} + \beta_{2i+1} = \frac{2}{(2+2i)(3+2i)} + \frac{2}{(3+2i)(4+2i)} = \frac{1}{(1+i)(2+i)} = \alpha_i \quad \text{for } i \in \mathbb{Z}^+. \tag{4.9}$$

It now follows from (2.5) that

$$c(u,v) = \sum_{i=0}^{\infty} \sum_{j=0}^{\infty} p_{ij} f_i(u) g_j(v), \ u,v \in (0,1) \tag{4.10}$$

is a copula density where $f_i(u) = \dfrac{(1-u)u^i}{\alpha_i}$ and $g_j(v) = \dfrac{(j+1)(1-v)^2 v^j}{\beta_j}$ for $i,j \in \mathbb{Z}^+$, $u,v \in (0,1)$. Using (4.8), one obtains, after some tedious but straightforward calculations, that

$$c(u,v) = \frac{2(1-u)(1-v)^2 \left(1 + 2v + 5uv^2 + 4uv^3\right)}{\left(1 - uv^2\right)^4}, \ u,v \in (0,1) \tag{4.11}$$

which obviously is asymmetric.

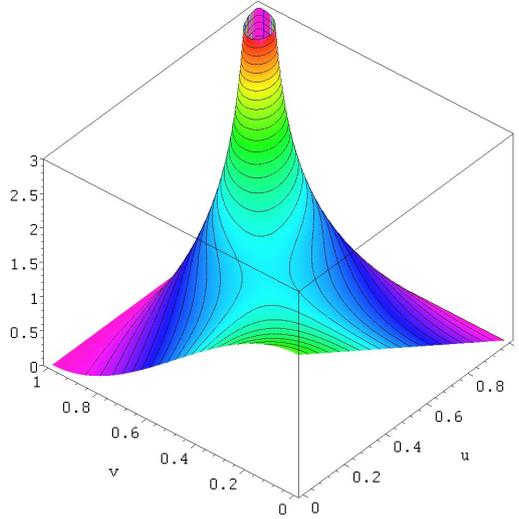

plot of $c(u,v)$

The corresponding copula $C$ can again be calculated explicitly, giving

$$C(x,y) = \frac{xy}{\left(1 - xy^2\right)^2} \left(2 - x - 2xy^3 + xy^4 + x^2 y^3 - 2y^2 + y^3\right), \ x,y \in (0,1). \tag{4.12}$$

This copula has a coefficient of upper tail dependence

$$\lambda_U = \frac{5}{9} \tag{4.13}$$

which is between the coefficients of upper tail dependence for the symmetric case with $\beta = 1$ and $\beta = 2$, cf. the final table in Example 2.

**Remark 1:** Negative binomial copulas (see Examples 2 and 5) can easily be simulated through the alternative representation formula (2.5) involving mixed Beta distributions here. Poisson copulas can be simulated using the transformation $z \mapsto 1 - e^{-z}$ applied to Gamma distributed random variables Z with a random shape parameter $\alpha$, where $\alpha - 1$ is generated by the geometric distribution shown in (3.15), and scale parameter $1 + \gamma$.

**Remark 2:** For practical applications in quantitative risk management, it seems reasonable to fit the required probabilities $\left[p_{ij}\right]_{i,j \in \mathbb{Z}^+}$ to empirical data via their empirical copula, for instance as was proposed in PFEIFER, STRASSBURGER AND PHILIPPS (2009). In the particular case of Bernstein copulas (see Example 1) such a procedure can be very easily implemented, even in higher dimensions (cf. COTTIN AND PFEIFER (2014)).

As a practical exercise, we refer to Example 4.2 in COTTIN AND PFEIFER (2014) where the empirical copula from an original data set was fitted to a general Bernstein copula. The following two graphs show the scatter plot from the empirical copula (big red dots) superimposed by 1000 simulated points of that Bernstein copula (left) and of a negative binomial copula of type (3.11), with $\beta = 5$.

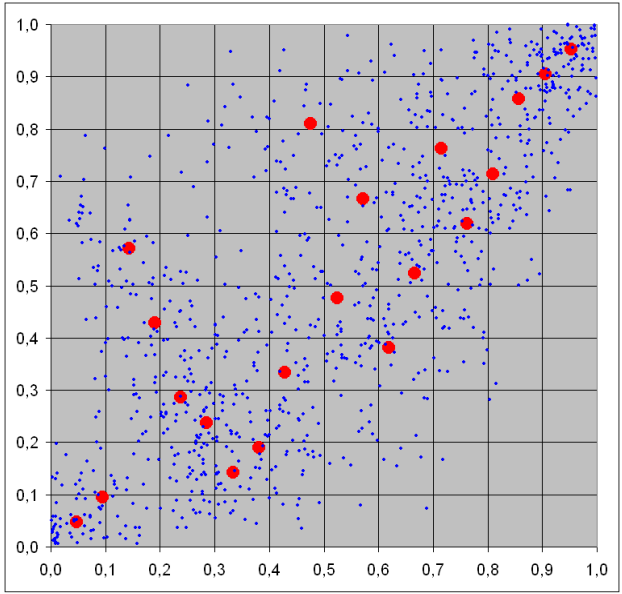
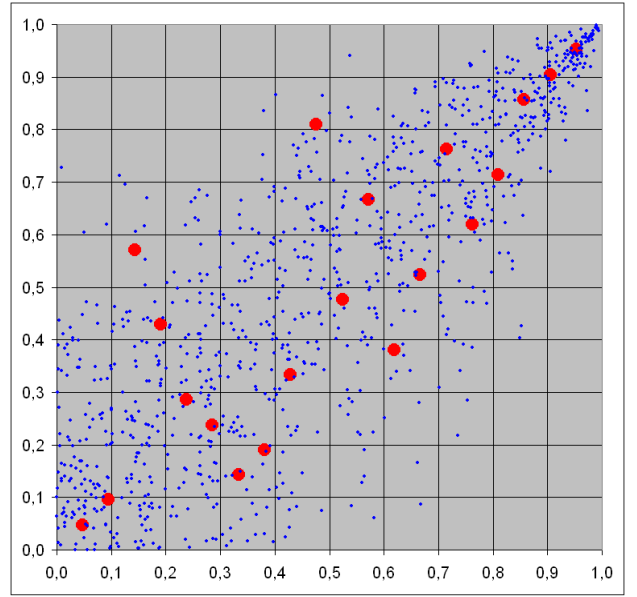

Bernstein copula fit            negative binomial copula fit

As can be seen nicely, the Bernstein copula represents the local asymmetry of the empirical copula better, but shows no tail dependence, as the negative binomial copula does.

The fit to the negative binomial copula was, for the sake of simplicity, performed by a numerical match between the theoretical correlation for the negative binomial copula and the correlation of the empirical copula, which is 0.815. Note that the theoretical correlation $\rho(\beta)$ for the negative binomial copula of type (3.11) can be explicitly calculated as

$$\rho(\beta) = 12\beta \left( \sum_{i=0}^{\infty} \frac{(i+1)^2}{(\beta+i)(\beta+i+1)(\beta+i+2)^2} \right) - 3 = 3\beta \left( 2(\beta+1)^2 \Psi(1,\beta+2) - 2\beta - 1 \right) \quad (4.14)$$

where $\Psi(1,z)$ denotes the first derivative of the digamma function, or $\Psi(1,z) = \dfrac{d^2}{dz^2}\ln\Gamma(z),\ z>0$.

| $\beta$ | 1 | 2 | 3 | 4 | 5 | 6 | 7 |
|---|---|---|---|---|---|---|---|
| $\rho(\beta)$ | 0.4784 | 0.6529 | 0.7410 | 0.7937 | 0.8288 | 0.8537 | 0.8723 |

For the sake of completeness, we finally show a comparison between the Bernstein copula fit and a Poisson copula fit with parameter $\gamma = 6$. The empirical correlation for the Poisson copula here is 0.814.

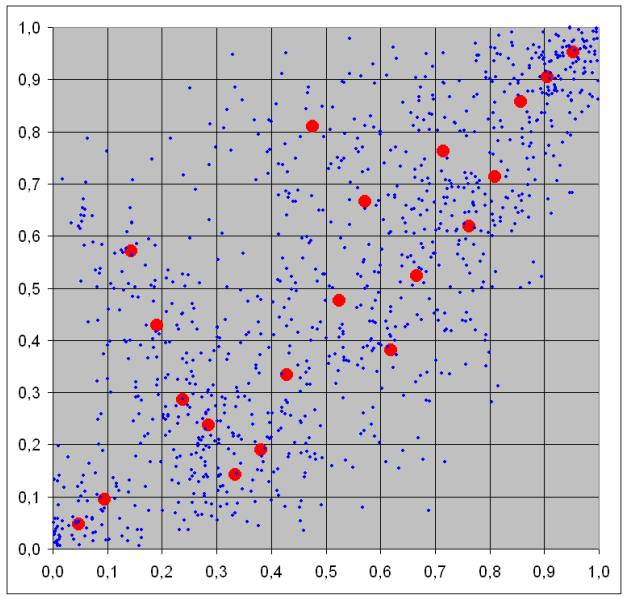
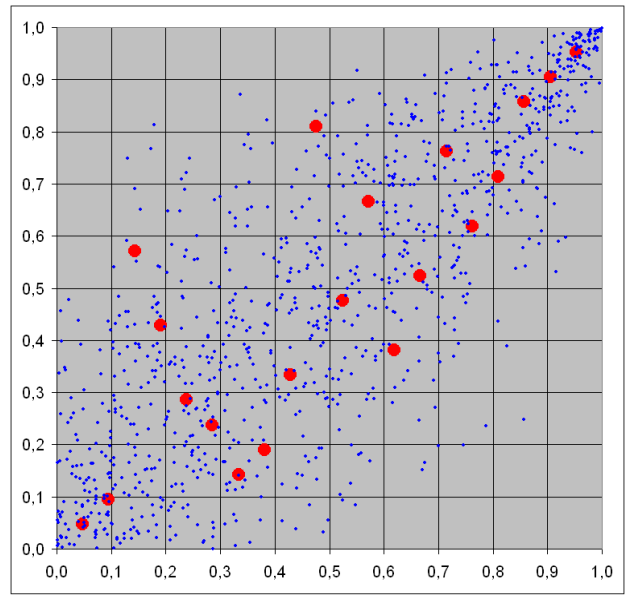

Bernstein copula fit    Poisson copula fit

Note that although the empirical plot for the Poisson copula might suggest some tail dependence here this is actually not true in the light of (3.19).

More sophisticated fitting procedures - including asymmetric cases – also in higher dimensions will be investigated elsewhere.

It should be finally pointed out that copula constructions as presented in this paper will have a major impact in the construction of Internal Models under the new Solvency II insurance supervising regime in Europe (see e.g. HUMMEL AND MÄRKERT (2011) or SANDSTRØM (2011), Chapter 13).

**5. Appendix.**

**Proof of Lemma 1.** We will show by induction the equality of the following two expressions:

$$K(\beta,z) := \sum_{i=0}^{\infty} \binom{\beta+i-1}{i}\binom{\beta+i+1}{i} z^i \ = \ k(\beta,z) := \frac{1}{(1-z)^{2\beta+1}} \sum_{i=0}^{\beta-1} \binom{\beta-1}{i}\binom{\beta+1}{i} z^i \quad \text{for } 0<z<1. \quad (5.1)$$

First notice that we have, for $\beta \in \mathbb{N}$ and $0<z<1$,

$$\frac{\partial K(\beta,z)}{\partial z} = \sum_{i=1}^{\infty} i \binom{\beta+i-1}{i}\binom{\beta+i+1}{i} z^{i-1} \text{ and } \frac{\partial^2 K(\beta,z)}{\partial z^2} = \sum_{i=2}^{\infty} i(i-1)\binom{\beta+i-1}{i}\binom{\beta+i+1}{i} z^{i-2} \quad (5.2)$$

from which we can conclude the relation

$$\frac{\partial^2 K(\beta,z)}{\partial z^2} = \frac{1}{z}\left(\beta(\beta+2)K(\beta+1,z) - \frac{\partial K(\beta,z)}{\partial z}\right) \text{ or } K(\beta+1,z) = \frac{z\frac{\partial^2 K(\beta,z)}{\partial z^2} + \frac{\partial K(\beta,z)}{\partial z}}{\beta(\beta+2)}. \quad (5.3)$$

A similar, but more elaborate calculation shows that the latter equality remains valid if $K(\beta,z)$ is replaced by $k(\beta,z)$:

$$k(\beta+1,z) = \frac{z\frac{\partial^2 k(\beta,z)}{\partial z^2} + \frac{\partial k(\beta,z)}{\partial z}}{\beta(\beta+2)}. \quad (5.4)$$

In the first step of the induction, for $\beta = 1$, we have

$$K(1,z) = \sum_{i=0}^{\infty} \frac{(i+1)(i+2)}{2} z^i = \sum_{j=2}^{\infty} \frac{j(j-1)}{2} z^{j-2} = \frac{1}{2} h''(z) = \frac{1}{(1-z)^3} = k(1,z) \quad (5.5)$$

with $h(z) := \sum_{i=0}^{\infty} z^i = \frac{1}{1-z}$ for $|z| < 1$. For the second step, assume that relation (5.01) holds for some $\beta \in \mathbb{N}$. Then it follows by (5.3) and (5.4) that

$$K(\beta+1,z) = \frac{z\frac{\partial^2 K(\beta,z)}{\partial z^2} + \frac{\partial K(\beta,z)}{\partial z}}{\beta(\beta+2)} = \frac{z\frac{\partial^2 k(\beta,z)}{\partial z^2} + \frac{\partial k(\beta,z)}{\partial z}}{\beta(\beta+2)} = k(\beta+1,z) \quad (5.6)$$

which finishes the proof. ◆

**Proof of Lemma 2.** First, note that for $\beta \in \mathbb{N}$,

$$\sum_{i=0}^{\beta-1} \binom{\beta-1}{i}\binom{\beta+1}{i} = \sum_{i=0}^{\beta-1} \binom{\beta-1}{\beta-1-i}\binom{\beta+1}{i} = \binom{2\beta}{\beta-1} \quad (5.7)$$

which is a special case of Vandermonde's identity. This in turn implies

$$(\beta+1)\sum_{i=0}^{\beta-1}\binom{\beta-1}{i}\binom{\beta+1}{i} = (\beta+1)\binom{2\beta}{\beta-1} = \frac{2\Gamma(2\beta)}{\Gamma^2(\beta)}. \quad (5.8)$$

Now, in the light of Lemma 1, we obtain

$$\lambda_U(\beta) = \lim_{h \downarrow 0} \frac{\int_{1-h}^{1}\int_{1-h}^{1} c_\beta(u,v)\,du\,dv}{h} = (\beta+1)\sum_{i=0}^{\beta-1}\binom{\beta-1}{i}\binom{\beta+1}{i}\lim_{h \downarrow 0} \frac{\int_{1-h}^{1}\int_{1-h}^{1} \frac{(1-u)^\beta(1-v)^\beta}{(1-uv)^{2\beta+1}}(uv)^i\,du\,dv}{h}. \quad (5.9)$$

To evaluate the last integral, we substitute $s = 1-u$, $w = 1-v$ and get

$$I(\beta,h,i) := \int_{1-h}^{1}\int_{1-h}^{1} \frac{(1-u)^\beta (1-v)^\beta}{(1-uv)^{2\beta+1}} (uv)^i \, du \, dv = \int_0^h \int_0^h \frac{s^\beta w^\beta}{(s+w-sw)^{2\beta+1}} (1-s)^i (1-w)^i \, ds \, dw. \tag{5.10}$$

In a further step, substituting $s = hx$, $w = hy$, we obtain

$$I(\beta,h,i) = h \int_0^1 \int_0^1 \frac{x^\beta y^\beta}{(x+y-hxy)^{2\beta+1}} (1-hx)^i (1-hy)^i \, dx \, dy, \tag{5.11}$$

giving

$$\lambda_U(\beta) = (\beta+1)\sum_{i=0}^{\beta-1}\binom{\beta-1}{i}\binom{\beta+1}{i}\lim_{h\downarrow 0}\frac{I(\beta,h,i)}{h} = (\beta+1)\sum_{i=0}^{\beta-1}\binom{\beta-1}{i}\binom{\beta+1}{i}\int_0^1\int_0^1 \frac{x^\beta y^\beta}{(x+y)^{2\beta+1}}\,dx\,dy$$

$$= \frac{2\Gamma(2\beta)}{\Gamma^2(\beta)} \cdot \int_0^1\int_0^1 \frac{x^\beta y^\beta}{(x+y)^{2\beta+1}}\,dx\,dy. \tag{5.12}$$

It remains to evaluate the integral term in the expression above. Therefore, we consider the one-to-one map $g : T \to (0,1)^2 : (u,v) \mapsto (uv, (1-u)v)$ for $T := \left\{(u,v) \in \mathbb{R}^2 \mid 0 < u < 1, 0 < v < \min\left(\frac{1}{u}, \frac{1}{1-u}\right)\right\}$. (Note that with $(x,y) := g(u,v)$, we have $u = \frac{x}{x+y}$, $v = x+y$ for $(x,y) \in (0,1)^2$.) By the substitution formula for multiple integrals, we now obtain, putting $f(x,y) := \frac{x^\beta y^\beta}{(x+y)^{2\beta+1}}$, and observing that for the determinant of the Jacobian, we have here $\det \Delta g(u,v) = v$,

$$\int_0^1\int_0^1 \frac{x^\beta y^\beta}{(x+y)^{2\beta+1}}\,dx\,dy = \iint_{g(T)} f(x,y)\,dx\,dy = \iint_T f(g(u,v)) \cdot |\det \Delta g(u,v)|\,du\,dv$$

$$= \iint_T u^\beta (1-u)^\beta \, du\, dv = \int_0^1 \min\left(\frac{1}{u},\frac{1}{1-u}\right) \cdot u^\beta (1-u)^\beta \, du = 2\int_0^{1/2} u^\beta (1-u)^{\beta-1}\,du \tag{5.13}$$

which proves the first line in (3.13). For the first equality in the second line, note that, by symmetry and the substitution $v = 2u$,

$$1 - \frac{4\Gamma(2\beta)}{\Gamma^2(\beta)} \cdot \int_0^{1/2} u^\beta (1-u)^{\beta-1}\,du = 2\frac{\Gamma(2\beta)}{\Gamma^2(\beta)} \cdot \left[\int_0^{1/2} u^{\beta-1}(1-u)^{\beta-1}\,du - \int_0^{1/2} 2u^\beta (1-u)^{\beta-1}\,du\right]$$

$$= 2\frac{\Gamma(2\beta)}{\Gamma^2(\beta)} \cdot \int_0^{1/2} u^{\beta-1}(1-u)^{\beta-1}(1-2u)\,du = \frac{\Gamma(2\beta)}{\Gamma^2(\beta)4^{\beta-1}} \cdot \int_0^1 (2-v)^{\beta-1} v^{\beta-1}(1-v)\,du$$

$$= \frac{\Gamma(2\beta)}{\Gamma^2(\beta)4^{\beta-1}} \cdot \left[\frac{(2z-z^2)z^{\beta-1}(2-z)^{\beta-1}}{2\beta}\right]_0^1 = \frac{\Gamma(2\beta)}{2\beta\Gamma^2(\beta)4^{\beta-1}} = \frac{(2\beta-1)!}{2\beta(\beta-1)!^2\,4^{\beta-1}} = \frac{\binom{2\beta}{\beta}}{4^\beta} \tag{5.14}$$

Which proves the first equality in the second line of (3.13). The asymptotic expansion follows by Stirling's formula. ◆

**Proof of Lemma 3.** Define $g_n(c) := \int_0^\infty \frac{(1-e^{-x})^n}{x} e^{-cx} dx$ for $c > 0$. Note that $f_n(x) := \frac{(1-e^{-x})^n}{x}$ for $x > 0$ is bounded by 1 for all $n \in \mathbb{N}$. We can therefore apply the dominated convergence theorem where appropriate. Now

$$g_n'(c) := -\int_0^\infty \frac{(1-e^{-x})^n}{x} \cdot x \cdot e^{-cx} dx = -\int_0^\infty (1-e^{-x})^n \cdot e^{-cx} dx = \int_0^\infty \sum_{j=0}^n \binom{n}{j} (-1)^{j+1} e^{-(j+c)x} dx$$

$$= \sum_{j=0}^n \binom{n}{j} (-1)^{j+1} \int_0^\infty e^{-(j+c)x} dx = \sum_{j=0}^n \binom{n}{j} (-1)^{j+1} \left( -\frac{1}{j+c} e^{-(j+c)x} \Big|_0^\infty \right) = \sum_{j=0}^n \binom{n}{j} (-1)^{j+1} \frac{1}{j+c} \quad (5.15)$$

for $c > 0$. Let further

$$h_n(c) := \sum_{j=0}^n \binom{n}{j} (-1)^{j+1} \left( \ln(j+c) - \ln(c) \right) \text{ for } c > 0. \quad (5.16)$$

Then

$$h_n'(c) = \sum_{j=0}^n \binom{n}{j} (-1)^{j+1} \frac{d}{dc} \left( \ln(j+c) - \ln(c) \right) = \sum_{j=0}^n \binom{n}{j} (-1)^{j+1} \left( \frac{1}{j+c} - \frac{1}{c} \right)$$

$$= \sum_{j=0}^n \binom{n}{j} (-1)^{j+1} \frac{1}{j+c} + \frac{1}{c} \sum_{j=0}^n \binom{n}{j} (-1)^j = \sum_{j=0}^n \binom{n}{j} (-1)^{j+1} \frac{1}{j+c} \quad (5.17)$$

since $0 = (1-1)^n = \sum_{j=0}^n \binom{n}{j} (-1)^j$. This implies $g_n' = h_n'$ and hence $g_n(c) = h_n(c) + K_n$ or equivalently, $K_n = g_n(c) - h_n(c)$ for all $c > 0$, for some constant $K_n \in \mathbb{R}$. But then also

$$K_n = \lim_{c \to \infty} g_n(c) - \lim_{c \to \infty} h_n(c) = \lim_{c \to \infty} \left( \int_0^\infty \frac{(1-e^{-x})^n}{x} e^{-cx} dx \right) - \lim_{c \to \infty} \sum_{j=0}^n \binom{n}{j} (-1)^{j+1} \left( \ln(j+c) - \ln(c) \right)$$

$$= \int_0^\infty \frac{(1-e^{-x})^n}{x} \lim_{c \to \infty} \left( e^{-cx} \right) dx - \sum_{j=0}^n \binom{n}{j} (-1)^{j+1} \lim_{c \to \infty} \left( \ln\left(1 + \frac{j}{c}\right) \right) = \int_0^\infty 0 \, dx - \sum_{j=0}^n 0 = 0 \quad (5.18)$$

for all $n \in \mathbb{N}$. Hence $g_n = h_n$ for all $n \in \mathbb{N}$, which proves the Lemma. ◆

**Proof of Lemma 4.** Substitute $s = 1 - t$. Then (3.26) is equivalent to

$$\lim_{s \downarrow 0} \frac{1}{s} \int_{1-s}^{1} \frac{L(u \cdot (1-s))}{L(u)} du = \lim_{s \downarrow 0} \frac{1}{s} \int_{0}^{s} \frac{L((1-w) \cdot (1-s))}{L(1-w)} dw = 1, \qquad (5.19)$$

with the substitution $u = 1 - w$. This means that we have to show that

$$\lim_{s \downarrow 0} \frac{1}{s} \int_{0}^{s} \frac{\ln(w + s - ws)}{\ln(w)} dw = 1. \qquad (5.20)$$

Define

$$F(w, s) := \frac{\ln(w + s)}{\ln(w)}, \quad G(w, s) := \frac{\ln(w + s(1-s))}{\ln(w)}. \qquad (5.21)$$

Then

$$F(w, s) \leq \frac{\ln(w + s - ws)}{\ln(w)} \leq G(w, s) \text{ for } 0 < w \leq s \qquad (5.22)$$

(note that $\ln(w) < 0$ for $0 < w < 1$). Now for $0 < s < 1$,

$$0 \leq \int_{0}^{s} G(w, s) - F(w, s) \, dw \leq \int_{0}^{s} \frac{-\ln\left(1 - \frac{s^2}{w+s}\right)}{-\ln(w)} dw \leq -\ln(1-s) \int_{0}^{s} \frac{1}{-\ln(w)} dw \leq \frac{s \ln(1-s)}{\ln(s)} \qquad (5.23)$$

with the limit

$$0 \leq \lim_{s \downarrow 0} \frac{1}{s} \int_{0}^{s} G(w, s) - F(w, s) \, dw \leq \lim_{s \downarrow 0} \frac{\ln(1-s)}{\ln(s)} = 0. \qquad (5.24)$$

Hence it suffices to prove

$$\lim_{s \downarrow 0} \frac{1}{s} \int_{0}^{s} F(w, s) \, dw = \lim_{s \downarrow 0} \frac{1}{s} \int_{0}^{s} \frac{-\ln(w+s)}{-\ln(w)} dw \stackrel{!}{=} 1. \qquad (5.25)$$

By the substitution $x = -\ln(w)$ we obtain the equivalent expression

$$\lim_{s \downarrow 0} \frac{1}{s} \int_{-\ln(s)}^{\infty} \frac{-\ln(e^{-x} + s)}{x} e^{-x} dx \stackrel{!}{=} 1. \qquad (5.26)$$

Note that

$$\int_{-\ln(s)}^{\infty} \frac{-\ln(e^{-x}+s)}{x} e^{-x}\,dx = \int_{-\ln(s)}^{\infty} \frac{-\ln(e^{-x}(1+se^{-x}))}{x} e^{-x}\,dx = \int_{-\ln(s)}^{\infty} \frac{x-\ln(1+se^{-x})}{x} e^{-x}\,dx$$

$$= \int_{-\ln(s)}^{\infty} e^{-x}\,dx - \int_{-\ln(s)}^{\infty} \frac{\ln(1+se^{-x})}{x} e^{-x}\,dx = s - \int_{-\ln(s)}^{\infty} \frac{\ln(1+se^{-x})}{x} e^{-x}\,dx. \qquad (5.27)$$

Hence it suffices to prove

$$\lim_{s\downarrow 0} \frac{1}{s} \int_{-\ln(s)}^{\infty} \frac{\ln(1+se^{-x})}{x} e^{-x}\,dx \overset{!}{=} 0. \qquad (5.28)$$

With the substitution $s = e^{-T}$, this is equivalent to

$$\lim_{T\uparrow\infty} e^{T} \int_{T}^{\infty} \frac{\ln(1+e^{x-T})}{x} e^{-x}\,dx \overset{!}{=} 0. \qquad (5.29)$$

Substituting finally $y = x - T$, this means

$$\lim_{T\uparrow\infty} e^{T} \int_{0}^{\infty} \frac{\ln(1+e^{y})}{y+T} e^{-(y+T)}\,dy = \lim_{T\uparrow\infty} \int_{0}^{\infty} \frac{\ln(1+e^{y})}{y+T} e^{-y}\,dy \overset{!}{=} 0. \qquad (5.30)$$

But this is now evident due to

$$0 \le \lim_{T\uparrow\infty} \int_{0}^{\infty} \frac{\ln(1+e^{y})}{y+T} e^{-y}\,dy \le \lim_{T\uparrow\infty} \int_{0}^{\infty} \frac{y+1}{y+T} e^{-y}\,dy = \int_{0}^{\infty} \lim_{T\uparrow\infty} \left( \frac{y+1}{y+T} e^{-y} \right) dy = 0 \qquad (5.31)$$

by Lebesgue's dominated convergence theorem. (For $T \ge 1$, an integrable majorant is given by $e^{-y}$.)

This proves Lemma 4. ◆

**Proof of Lemma 5.** First notice that by the relation $L(u) = -\ln(1-u) = \ln\left(\frac{1}{1-u}\right) \le \frac{1}{1-u} - 1 = \frac{u}{1-u}$ for $0 < u < 1$, we obtain

$$\alpha_i = \int_{0}^{1} \frac{u^i}{i\cdot L(u)}\,du \ge \frac{1}{i} \int_{0}^{1} \frac{u^i}{u}(1-u)\,du = \frac{1}{i} \int_{0}^{1} u^{i-1}(1-u)\,du = \frac{1}{i^2(i+1)} \quad \text{for all } i \in \mathbb{N}. \qquad (5.32)$$

Now

$$K(t) := \frac{1}{1-t} \int_{t}^{1} \int_{t}^{1} \sum_{i=1}^{\infty} \frac{(uv)^i}{\alpha_i i^2 L(u)L(v)}\,du\,dv = \frac{1}{1-t} \int_{t}^{1} \left\{ \sum_{i=1}^{\infty} \frac{v^i}{\alpha_i i^2 L(v)} \int_{t}^{1} \frac{u^i}{L(u)}\,du \right\} dv \qquad (5.33)$$

by Lebesgue's dominated convergence theorem since for fixed $v \in (0,1)$, by (5.32),

$$\sum_{i=1}^{\infty} \frac{(uv)^i}{\alpha_i i^2 L(u) L(v)} \leq \sum_{i=1}^{\infty} (i+1) \frac{u^i}{L(u)} \cdot \frac{v^i}{L(v)} \leq \sum_{i=1}^{\infty} (i+1)(uv)^{i-1} = \frac{2+(uv)^2 - 3uv}{(1-uv)^3} \leq \frac{3}{(1-uv)^3}, \quad (5.34)$$

the r.h.s being integrable w.r.t. $u$ with value $\int_0^1 \frac{3}{(1-uv)^3} du = \frac{3(2-v)}{2(1-v)^2}$. Now for $0 < t < 1$, we have

$$\int_0^t \frac{u^i}{L(u)} du = t \int_0^1 \frac{v^i t^i}{L(vt)} dv \geq t^{i+1} \int_0^1 \frac{v^i}{L(v)} dv \quad (5.35)$$

and hence

$$\int_t^1 \frac{u^i}{L(u)} du = \int_0^1 \frac{u^i}{L(u)} du - \int_0^t \frac{u^i}{L(u)} du \leq \left(1 - t^{i+1}\right) \int_0^1 \frac{v^i}{L(v)} dv = \left(1 - t^{i+1}\right) i \alpha_i \quad (5.36)$$

for all $i \in \mathbb{N}$. Thus we get, from (5.33),

$$K(t) = \frac{1}{1-t} \int_t^1 \left\{ \sum_{i=1}^{\infty} \frac{v^i}{\alpha_i i^2 L(v)} \int_t^1 \frac{u^i}{L(u)} du \right\} dv \leq \frac{1}{1-t} \int_t^1 \sum_{i=1}^{\infty} \frac{v^i}{iL(v)} \left(1 - t^{i+1}\right) dv$$

$$= \frac{1}{1-t} \int_t^1 \frac{1}{L(v)} \left\{ \sum_{i=1}^{\infty} \frac{v^i}{i} - t \sum_{i=1}^{\infty} \frac{(vt)^i}{i} \right\} dv = \frac{1}{1-t} \int_t^1 \frac{L(v) - tL(vt)}{L(v)} dv = 1 - \frac{t}{1-t} \int_t^1 \frac{L(vt)}{L(v)} dv, \quad (5.37)$$

which proves relation (3.27). From Lemma 4 we thus obtain the final result

$$0 \leq \lambda_U = \lim_{t \uparrow 1} K(t) \leq 1 - 1 = 0 \text{ and hence } \lambda_U = 0, \quad (5.38)$$

which proves Lemma 5. ◆

**Acknowledgements.** We would like thank the referees for several helpful comments which improved the presentation of the paper substantially, and especially for pointing out the relationship to Catalan numbers in Lemma 2 and the corresponding references given in *The On-Line Encyclopedia of Integer Sequences*©.

**Obituary.** Sadly, Hervé Awoumlac Tsatedem has passed away in November 2015 at the early age of 31 due to an unforeseen heart attack.